\documentclass[12pt,a4paper,final]{iopart}

\usepackage{iopams}  
\usepackage{graphicx}
\usepackage[breaklinks=true,colorlinks=true,linkcolor=blue,urlcolor=blue,citecolor=blue]{hyperref}
\usepackage{braket}
\usepackage{xcolor}
\usepackage{sidecap}
\usepackage{mathrsfs}

\begin{document}

\title[]{Short wavelength limit of the dynamic Matsubara local field correction}

\author{Tobias Dornheim}
\address{Center for Advanced Systems Understanding (CASUS), 03581 G\"orlitz, Germany}
\address{Helmholtz-Zentrum Dresden-Rossendorf (HZDR), 01328 Dresden, Germany}
\ead{t.dornheim@hzdr.de}

\author{Panagiotis Tolias}
\address{Space and Plasma Physics, Royal Institute of Technology (KTH), Stockholm SE-100 44, Sweden}

\author{Zhandos A.~Moldabekov}
\address{Center for Advanced Systems Understanding (CASUS), 03581 G\"orlitz, Germany}
\address{Helmholtz-Zentrum Dresden-Rossendorf (HZDR), 01328 Dresden, Germany}

\author{Jan Vorberger}
\address{Helmholtz-Zentrum Dresden-Rossendorf (HZDR), 01328 Dresden, Germany}

\begin{abstract}
We investigate the short wavelength limit of the dynamic Matsubara local field correction $\widetilde{G}(\mathbf{q},z_l)$ of the uniform electron gas based on direct \emph{ab initio} path integral Monte Carlo (PIMC) results over an unprecedented range of  wavenumbers, $q\lesssim20q_\textnormal{F}$, where $q_\textnormal{F}$ is the Fermi wavenumber. We find excellent agreement with the analytically derived asymptotic limit by Hou \emph{et al.}~[\textit{Phys.~Rev.~B}~\textbf{106}, L081126 (2022)] for the static local field correction and empirically confirm the independence of the short wavelength limit with respect to the Matsubara frequency $z_l$. In the warm dense matter regime, we find that the onset of the quantum tail in the static local field correction closely coincides with the onset of the algebraic tail in the momentum distribution function and the corresponding empirical criterion reported by Hunger \emph{et al.}~[\textit{Phys.~Rev.~E} \textbf{103}, 053204 (2021)]. In the strongly coupled electron liquid regime, our calculations reveal a more complicated non-monotonic convergence towards the $q\to\infty$ limit that is shaped by the spatial structure in the system. 
We expect our results to be of broad interest for a number of fields including the study of matter under extreme conditions, the development of improved dielectric theories, and the construction of advanced exchange--correlation functionals for thermal density functional theory.
\end{abstract}
\vspace{2pc}

\section{Introduction\label{sec:introduction}}

The uniform electron gas (UEG)~\cite{quantum_theory,loos,review}, also known as jellium and quantum one-component plasma in the literature, constitutes the archetypal system of interacting electrons. Indeed, due to the simplified treatment of the ions as a homogeneous neutralizing background, the UEG allows one to exclusively focus on electronic effects. Consequently, the accurate parametrization of UEG properties~\cite{Perdew_Wang_PRB_1992,Perdew_Zunger_PRB_1981,vwn,cdop,Gori-Giorgi_PRB_2000,Gori-Giorgi_PRB_2002,Gori-Giorgi_Momentum_2002,ksdt,Brown_PRB_2013,groth_prl,dornheim_ML,Dornheim_PRB_ESA_2021,Karasiev_status_2019} based on highly accurate quantum Monte Carlo (QMC) simulations both at ambient conditions~\cite{Foulkes_RMP_2001,Ceperley_Alder_PRL_1980,Ortiz_PRB_1994,Ortiz_PRL_1999,moroni,moroni2} and at finite temperatures~\cite{Brown_PRL_2013,dornheim_prl,Schoof_PRL_2015,Malone_PRL_2016,Hunger_PRE_2021,Militzer_Pollock_PRL_2002} has been of paramount importance for a host of practical applications, most notably as an indispensable input for density functional theory (DFT) simulations of real materials~\cite{Jones_RMP_2015,bonitz2024principles}.

A particularly important class of information is given by the linear response of the UEG, which is typically expressed in terms of the dynamic density response function~\cite{kugler1}
\begin{eqnarray}\label{eq:define_G}
    \chi(\mathbf{q},\omega) = \frac{ \chi_0(\mathbf{q},\omega) }{ 1 - 4\pi/q^2\left[1 - G(\mathbf{q},\omega)\right]\chi_0(\mathbf{q},\omega)}\quad ,
\end{eqnarray}
where $\mathbf{q}$ and $\omega$ are the wave vector and frequency of an external harmonic perturbation~\cite{Dornheim_review}. Here $\chi_0(\mathbf{q},\omega)$ denotes the Lindhard function that describes the density response of a noninteracting Fermi gas, and the complete wave vector and frequency resolved information about electronic exchange--correlation (XC) effects is contained in the dynamic local field correction (LFC) $G(\mathbf{q},\omega)$. Hence, setting $G(\mathbf{q},\omega)\equiv0$ in Eq.~(\ref{eq:define_G}) leads to a description of the density response on the mean field level, which is commonly known as \emph{random phase approximation} (RPA) in the literature.
The LFC constitutes key input for a broad variety of applications, including the estimation of ionization potential depression~\cite{Zan_PRE_2021}, the construction of electronically screened effective potentials~\cite{ceperley_potential,Poole_PRR_2024}, and stopping power calculations~\cite{Cayzac2017}. Moreover, it is formally related to the XC-kernel
\begin{eqnarray}\label{eq:kernel}
    K_\textnormal{xc}(\mathbf{q},\omega) = - \frac{4\pi}{q^2}G(\mathbf{q},\omega)\ ,
\end{eqnarray}
which is a key property in linear-response time-dependent DFT (LR-TDDFT) simulations~\cite{book_Ullrich,Moldabekov_PRR_2023} and for the construction of advanced XC-functionals for equilibrium Kohn-Sham DFT~\cite{pribram,Patrick_JCP_2015,Lu_JCP_2024}. Finally, we note that approximate closure relations for the LFC are of central importance to so-called dielectric theories~\cite{stls_original,vs_original,IIT,dynamic_ii,stls,stls2,schweng,arora,Tanaka_CPP_2017,tanaka_hnc,dornheim_electron_liquid,castello2021classical,Tolias_JCP_2021,Tolias_JCP_2023,Tolias_PRB_2024}, which have been shown to be highly successful for the description of the UEG over a broad range of parameters.

The first highly accurate results for the LFC have been presented by Moroni \emph{et al.}~\cite{moroni,moroni2} based on ground-state QMC simulations in the static limit of $\omega=0$. These results were subsequently parametrized by Corradini \emph{et al.}~\cite{cdop}, who included the correct $q\to0$ and $q\to\infty$ limits, which are given by the compressibility sum-rule~\cite{quantum_theory,Tolias_PRB_2024} and a somewhat more complicated expression derived by Holas~\cite{holas_limit}, respectively.
Subsequently, Dornheim and co-workers~\cite{dornheim_ML,dornheim_HEDP,Dornheim_HEDP_2022,dornheim_electron_liquid} have presented extensive results for the static LFC at finite temperatures based on quasi-exact path integral Monte Carlo (PIMC) simulations. These efforts have focused on the \emph{warm dense matter} (WDM) regime, which is an extreme state that is characterized by the simultaneous presence of Coulomb correlations, partial quantum degeneracy, and strong thermal excitations~\cite{new_POP,wdm_book}.
Such WDM naturally occurs in a multitude of astrophysical objects such as giant planet interiors~\cite{Benuzzi_Mounaix_2014,wdm_book}
and white dwarf atmospheres~\cite{Kritcher_Nature_2020,Saumon_PhysReports_2022}, and is realized as a transient state e.g.~on the compression path of the fuel capsule and the ablator in inertial fusion energy applications~\cite{hu_ICF,Betti2016,Hurricane_RevModPhys_2023}.
In particular, these PIMC results have been used to construct a neural network representation of $G(\mathbf{q},0;r_s,\Theta)$ [where $r_s=d/a_\textnormal{B}$ is the Wigner-Seitz radius in units of the Bohr radius and $\Theta=k_\textnormal{B}T/E_\textnormal{F}$ the degeneracy temperature~\cite{Ott2018}, with the Fermi energy $E_\textnormal{F}$]
that covers the full WDM parameter regime and consistently includes the ground-state limit~\cite{dornheim_ML}. For completeness, we note that a semi-empirical extension is given by the \emph{effective static approximation} (ESA)~\cite{Dornheim_PRL_2020_ESA}, which has recently become available also as an analytical parametrization~\cite{Dornheim_PRB_ESA_2021}.

A first important limitation by these results is given by their restriction to $\omega\to0$. This is problematic for applications that aim to describe dynamic properties, such as LR-TDDFT calculations for the electronic dynamic structure factor $S_{ee}(\mathbf{q},\omega)$---the key observable in x-ray Thomson scattering (XRTS) experiments~\cite{siegfried_review,Schoerner_PRE_2023,gawne2024ultrahigh}, which constitute an important method of diagnostics for extreme states of matter~\cite{Gregori_PRE_2003,Tilo_Nature_2023,Dornheim_T_2022,Dornheim_T2_2022}.
This problem was partially solved in Refs.~\cite{dornheim_dynamic,dynamic_folgepaper,Hamann_PRB_2020} based on an analytic continuation of the imaginary-time density--density correlation function (ITCF) $F(\mathbf{q},\tau)=\braket{\hat{n}(\mathbf{q},\tau),\hat{n}(-\mathbf{q},0)}$, i.e., the numerical inversion of the well-known relation
\begin{eqnarray}
    F(\mathbf{q},\tau) = \int_{-\infty}^\infty \textnormal{d}\omega\ S_{ee}(\mathbf{q},\omega)\ e^{-\omega\tau}\ ;
\end{eqnarray}
note that we assume Hartree atomic units throughout unless indicated otherwise. While this inversion constitutes a notoriously difficult inverse problem~\cite{JARRELL1996133}, it was tamed for the special case of the UEG by including a number of analytically known exact constraints of $G(\mathbf{q},\omega)$~\cite{Dabrowski_PRB_1986} into a stochastic sampling procedure. This allowed Hamann \emph{et al.}~\cite{Hamann_PRB_2020} to show the first explicit results for the dynamic LFC of the UEG for a number of state points. In a subsequent proof-of-principle study, LeBlanc \emph{et al.}~\cite{LeBlanc_PRL_2022} presented a few additional results for low temperatures and moderate coupling, but readily available results for the dynamic LFC remain sparse.

A second limitation of previous results is given by the restricted range of available wave numbers $q=|\mathbf{q}|$. First, QMC simulations in a finite simulation cell of volume $\Omega=L^3$ are only available above the minimum wavenumber $q_\textnormal{min}=2\pi/L$. Second, while accessing large $q\gg q_\textnormal{F}$ [where $q_\textnormal{F}$ is the Fermi wavenumber~\cite{quantum_theory}] is, in principle, straightforward, the effect of the LFC onto the physical observable---$F(\mathbf{q},\tau)$ in the case of PIMC---rapidly decreases, making it difficult to resolve the former.
This problem was recently overcome by Hou \emph{et al.}~\cite{Hou_PRB_2022}, who have employed a novel variational diagrammatic Monte Carlo method to resolve the static XC-kernel $K_\textnormal{xc}(\mathbf{q},0)$ [and, hence, the static LFC $G(\mathbf{q},0)$] up to $q\sim10q_\textnormal{F}$ at moderate coupling, $r_s=1$, over a wide range of temperatures. 
Moreover, they have introduced a thermal generalization of the short wavelength asymptote by Holas~\cite{holas_limit}, which was fully confirmed by the new QMC results.

In the present work, we extend the existing body of results in two distinct ways. First, we carry out very extensive \emph{ab initio} PIMC simulations to resolve the static LFC over an unprecedented range of wavenumbers, $0.5q_\textnormal{F} \lesssim q \lesssim 20q_\textnormal{F}$. This allows us to validate the short wavelength asymptotic by Hou \emph{et al.}~\cite{Hou_PRB_2022}, covering a broad range of densities from warm dense matter ($r_s=2$) to the strongly coupled electron liquid~\cite{dornheim_electron_liquid} ($r_s=100$). In the former case, we find that the onset of the asymptotic nicely coincides with the onset of the algebraic tail in the momentum distribution function that has been studied by Hunger \emph{et al.}~\cite{Hunger_PRE_2021}; in the latter case, the situation is more complicated due to the integer harmonics of the \emph{roton-type feature}~\cite{Dornheim_Nature_2022,Trigger,Godfrin2012} in the spectrum of collective excitations~\cite{Dornheim_EPL_2024}.
Second, we go beyond the static limit of $\omega=0$ and investigate the full Matsubara density response function $\widetilde{\chi}(\mathbf{q},z_l)$~\cite{tolias2024fouriermatsubara,dornheim2024MatsubaraResponse,Dornheim_EPL_2024} and the corresponding dynamic Matsubara local field correction $\widetilde{G}(\mathbf{q},z_l)$, with $z_l=i2\pi l/\beta$ [and $-\infty,\dots,l,\dots,\infty$ an integer number] being the imaginary bosonic Matsubara frequencies.
This allows us to empirically validate Hou \emph{et al.'s}~\cite{Hou_PRB_2022} earlier prediction about the frequency-independence of the short wavelength limit, although the onset wavenumber increases with the frequency index $l$.

In addition to being interesting in their own right, our results further complete the understanding of the UEG, which is arguably the most important model system in physics and quantum chemistry~\cite{quantum_theory}. Moreover, understanding the asymptotic limits of $G(\mathbf{q},0)$ and $\widetilde{G}(\mathbf{q},z_l)$ is indispensable for the future construction of a four-point parametrization $\widetilde{G}(\mathbf{q},z_l;r_s,\Theta)$, which would be highly useful for a broad range of applications such as the construction of advanced XC-functionals for thermal DFT simulations~\cite{pribram}. The paper is organized as follows: In Sec.~\ref{sec:theory}, we provide the theoretical background, including our PIMC framework to the dynamic Matsubara density response in Sec.~\ref{sec:PIMC}, expressions for the short wavelength limit in Sec.~\ref{sec:short_wavelength_limit}, a brief discussion of the quantum tail in the momentum distribution function in Sec.~\ref{sec:tail}, and the high-frequency limit of the dynamic Matsubara LFC in Sec.~\ref{sec:high_frequency}.
Sec.~\ref{sec:results} covers our new simulation results, with Sec.~\ref{sec:WDM_results} and Sec.~\ref{sec:electron_liquid} being devoted to the warm dense matter and electron liquid regimes, respectively. The paper is concluded by a summary and outlook in Sec.~\ref{sec:summary}.

\section{Theory\label{sec:theory}}

\subsection{PIMC approach to the dynamic Matsubara density response\label{sec:PIMC}}

Having originally being introduced for the description of ultracold $^4$He~\cite{Fosdick_PR_1966,Jordan_PR_1968}, the \emph{ab initio} PIMC method has emerged as a powerful tool for the description of interacting quantum many-body systems at finite temperatures~\cite{cep}. The basic idea is to stochastically sample the thermal density matrix $\hat{\rho}=e^{-\beta\hat{H}}$ [where $\hat{H}$ is the Hamilton operator] evaluated in coordinate space, which, in principle, gives one access to all thermodynamic properties of a given system; more detailed introductions to PIMC have been presented in the literature~\cite{cep,boninsegni1}.
An additional obstacle for PIMC simulations of WDM, in general, and of the warm dense UEG, in particular, is given by the notorious fermion sign problem~\cite{dornheim_sign_problem,troyer}. It originates from the anti-symmetry of the fermionic density matrix under the exchange of particle coordinates and leads to an exponential increase in the required compute time with important system parameters such as the system size $N$ or the inverse temperature $\beta$. 
While a number of ideas to alleviate the sign problem have been suggested over the years, e.g.~Refs.~\cite{Ceperley1991,Brown_PRL_2013,DuBois,Schoof_CPP_2015,Schoof_PRL_2015,Dornheim_NJP_2015,Hirshberg_JCP_2020,Dornheim_JCP_2020,Xiong_JCP_2022,Xiong_PRE_2023,Dornheim_JCP_2023,Chin_PRE_2015,Blunt_PRB_2014,Malone_PRL_2016,Joonho_JCP_2021,Clark_PRB_2017,Egger_PRE_2000,Egger_PRL_1998,PhysRevB.63.235105}, the bulk of them does not readily give one access to the ITCF $F(\mathbf{q},\tau)$ at the present time. Therefore, we carry out direct PIMC simulations that are subject for the full sign problem; this makes our simulations computationally involved, but exact within the given Monte Carlo error bars.

Very recently, Tolias \emph{et al.}~\cite{tolias2024fouriermatsubara} have introduced an exact Fourier--Matsubara series representation of the ITCF, which is given by
\begin{eqnarray}\label{eq:FMITCF}
    F(\boldsymbol{q},\tau)=-\frac{1}{n\beta}\sum_{l=-\infty}^{+\infty}\widetilde{\chi}(\mathbf{q},z_l)e^{-z_l\tau}\,;
\end{eqnarray}
it constitutes the imaginary-time generalization of the usual Matsubara series of the static structure factor as $F(\boldsymbol{q},\tau=0)=S(\boldsymbol{q})$.
We note that the ITCF has attracted remarkable recent interest due to its value for the diagnostics of X-ray scattering experiments with matter under extreme conditions~\cite{Dornheim_T_2022,Dornheim_T2_2022,Schoerner_PRE_2023,dornheim2023xray,Dornheim_review,Dornheim_Science_2024,Dornheim_PTR_2022,Dornheim_insight_2022}. In the present work, we focus on the dynamic Matsubara density response, i.e., the coefficients in Eq.~(\ref{eq:FMITCF}) that can be computed from the ITCF in a straightforward way via
\begin{eqnarray}\label{eq:MDR}
    \widetilde{\chi}(\mathbf{q},z_l) = -2n\int_0^{\beta/2}\textnormal{d}\tau\ F(\mathbf{q},\tau)\ \textnormal{cos}\left(i z_l \tau\right)\,.
\end{eqnarray}
It is easy to see that Eq.~(\ref{eq:MDR}) constitutes the dynamic generalization of the imaginary-time version of the fluctuation--dissipation theorem~\cite{Dornheim_insight_2022}, which has been pivotal for the PIMC based investigation of a host of linear-response properties, e.g., Refs.~\cite{dornheim_ML,dynamic_folgepaper,dornheim_electron_liquid,dornheim_HEDP,Dornheim_HEDP_2022}. The corresponding dynamic Matsubara local field correction is then obtained by inverting Eq.~(\ref{eq:define_G}),
\begin{eqnarray}\label{eq:get_G}
    \widetilde{G}(\mathbf{q},z_l) = 1 + \frac{q^2}{4\pi}\left\{
\frac{1}{\widetilde{\chi}(\mathbf{q},z_l)} - \frac{1}{\widetilde{\chi}_0(\mathbf{q},z_l)}
    \right\}\ .
\end{eqnarray}
From Eq.~(\ref{eq:get_G}), it is directly evident that $\widetilde{G}(\mathbf{q},z_l)$ is given by the difference between two similar numbers for large $q$ as $\widetilde{\chi}(\mathbf{q},z_l)$ converges towards $\widetilde{\chi}_0(\mathbf{q},z_l)$, leading to increasing statistical error bars.

\subsection{Short wavelength limit\label{sec:short_wavelength_limit}}

In the ground-state limit of $\Theta=0$, the short wavelength limit of the static LFC of the UEG has been investigated by Niklasson with asymptotic quantum kinetic theory~\cite{NiklassonLimit}, by Holas with asymptotic linear response theory~\cite{holas_limit} and by Vignale with asymptotic diagrammatic many body theory~\cite{Vignale_PRB_1988}. Hou \emph{et al.}~\cite{Hou_PRB_2022} extended the analysis of Vignale to the finite temperature UEG utilizing Green's functions within the Matsubara formalism. Regardless of the UEG temperature, given that $\widetilde{G}(x\to\infty,l=0)=G(x\to\infty,0)$, the exact short wavelength behavior is
\begin{eqnarray}\label{eq:Gzeroshort}
   \widetilde{G}(x\to\infty,l=0)= \frac{1}{2}\pi\lambda{r}_{\mathrm{s}}T_{\mathrm{xc}}x^2+\mathcal{O}(x^0)\,,
\end{eqnarray}
with $x=q/q_\textnormal{F}$, $\lambda=1/(r_s q_{\mathrm{F}})=[4/(9\pi)]^{1/3}$, and $T_\textnormal{xc}=T-T_0$ being the XC-contribution to the kinetic energy in dimensionless units, i.e., divided by the Hartree energy.

Hou \emph{et al.}~\cite{Hou_PRB_2022} also pointed out that the above result can be generalized to arbitrary frequencies,  since the leading order term does not have a Matsubara frequency dependence. Therefore, 
\begin{eqnarray}\label{eq:Gzeroshortgen}
   \widetilde{G}(x\to\infty,l)= \frac{1}{2}\pi\lambda{r}_{\mathrm{s}}T_{\mathrm{xc}}x^2\,.
\end{eqnarray}
It should be emphasized that, in contrast to the ground state UEG for which $T_{\mathrm{xc}}>0$ for all $r_{\mathrm{s}}$~\cite{quantum_theory}, the XC contribution to the kinetic energy can obtain any sign for the finite temperature UEG~\cite{Hou_PRB_2022,Militzer_Pollock_PRL_2002,Hunger_PRE_2021,Dornheim_PRB_nk_2021}. For instance, at the Fermi temperature, $T_{\mathrm{xc}}<0$ for $r_{\mathrm{s}}\lesssim4$ and $T_{\mathrm{xc}}>0$ for $r_{\mathrm{s}}\gtrsim4$~\cite{Hunger_PRE_2021,Dornheim_PRB_nk_2021}.
This has interesting consequences for the ordering of $\widetilde{G}(\mathbf{q},z_l)$ with respect to $l$ at large $q$ in the warm dense matter regime as we demonstrate in Sec.~\ref{sec:WDM_results} below.

\subsection{Large momentum tail\label{sec:tail}}

It is well-known that the momentum distribution function $n(q)$ of an interacting quantum many-body system qualitatively deviates from the Fermi distribution function due to an algebraic tail at large momenta~\cite{Hoffmann_PRB_2013}.
For protons (as well as deuterium and tritium nuclei), this effect is expected to play an important role in the enhancement of fusion rates~\cite{Starostin2005}.
The quantum tail in the momentum distribution of the unpolarized UEG is given by~\cite{Hoffmann_PRB_2013,Hunger_PRE_2021}
\begin{eqnarray}\label{eq:momentum_tail}
  n^\infty(q) =  \lim_{q\to\infty}n(q) = \frac{8}{9\pi^2}\left(
\alpha r_s
    \right)^2 g(0) \left(
\frac{q_\textnormal{F}}{q}
    \right)^8\ .
\end{eqnarray}
Hunger \emph{et al.}~\cite{Hunger_PRE_2021} have found empirically based on highly accurate configuration PIMC results for $n(q)$ that the onset wavenumber $q_\textnormal{tail}$ of the quantum tail in the momentum distribution can be determined from the intersection of Eq.~(\ref{eq:momentum_tail}) with the Fermi distribution $n_0(q)$,
\begin{eqnarray}\label{eq:q_tail}
    n_0(q_\textnormal{tail}) = n^\infty(q_\textnormal{tail})\ ;
\end{eqnarray}
this procedure is illustrated in Fig.~\ref{fig:Momentum_rs2_theta2} below.
In the present work, we find that Eq.~(\ref{eq:q_tail}) also constitutes a very useful criterion to predict the onset of the short wavelength limit of the static LFC in the WDM regime, which might be an important observation for the future construction of a four-point parametrization $\widetilde{G}(\mathbf{q},z_l;r_s,\Theta)$.

\subsection{High-frequency asymptotic\label{sec:high_frequency}}

An additional useful limit of the dynamic Matsubara LFC is its high-frequency asymptotic $\widetilde{G}(\mathbf{q},\mathrm{i}\infty) = G(\mathbf{q},\infty)$.
It is given by~\cite{quantum_theory,kugler1,holas_limit,dornheim2024MatsubaraResponse}
\begin{eqnarray}\label{eq:Ginfty}
   G(\mathbf{q},\infty) = I(\mathbf{q}) - \frac{2q^2}{\omega_\textnormal{p}^2}T_\textnormal{xc}\,,
\end{eqnarray}
with $\omega_\textnormal{p}=\sqrt{3/r_s^3}$ being the plasma frequency and $I(\mathbf{q})$ the well-known Pathak-Vashishta functional~\cite{PathakVashishtaScheme,NiklassonLimit,SingwiTosi_Review,dornheim2024MatsubaraResponse} that is a functional of the static structure factor $S(\mathbf{q})=F(\mathbf{q},0)$.
A PIMC based analysis of Eq.~(\ref{eq:Ginfty}) in terms of the dynamic Matsubara density response function has been presented in the recent Ref.~\cite{dornheim2024MatsubaraResponse}.

\section{Results\label{sec:results}}

All PIMC results that are presented in this work have been obtained using the extended ensemble scheme~\cite{Dornheim_PRB_nk_2021} as it is implemented in the \texttt{ISHTAR} code~\cite{ISHTAR}; it is a canonical adaption of the seminal worm algorithm by Boninsegni \emph{et al.}~\cite{boninsegni1,boninsegni2}, and allows for an ergodic sampling in all permutation sectors~\cite{Dornheim_permutation_cycles}.
All simulation results are available online~\cite{repo} and can be used freely for other applications.

\subsection{Warm dense matter\label{sec:WDM_results}}

\begin{figure}\centering\includegraphics[width=0.45\textwidth]{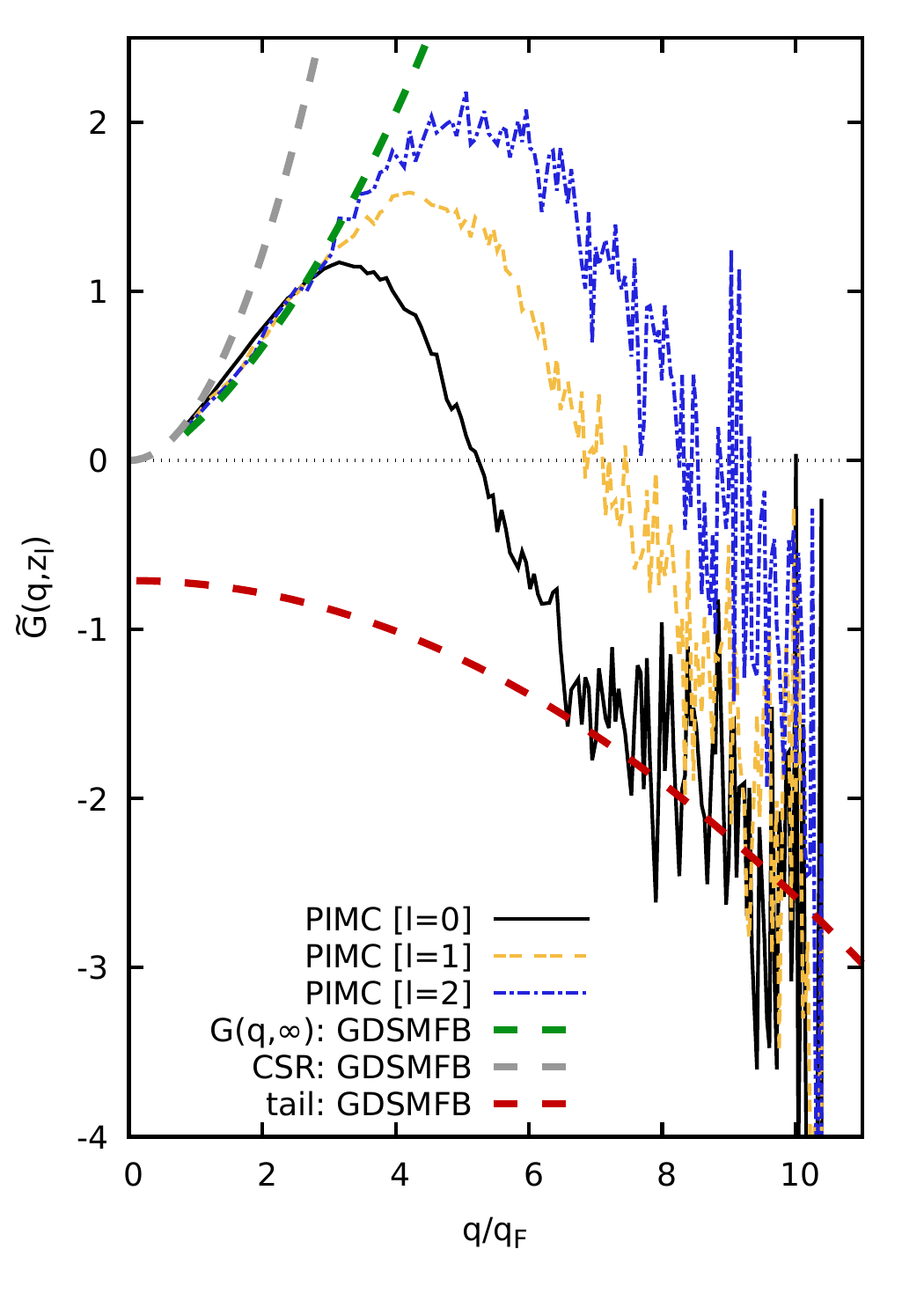}
\includegraphics[width=0.45\textwidth]{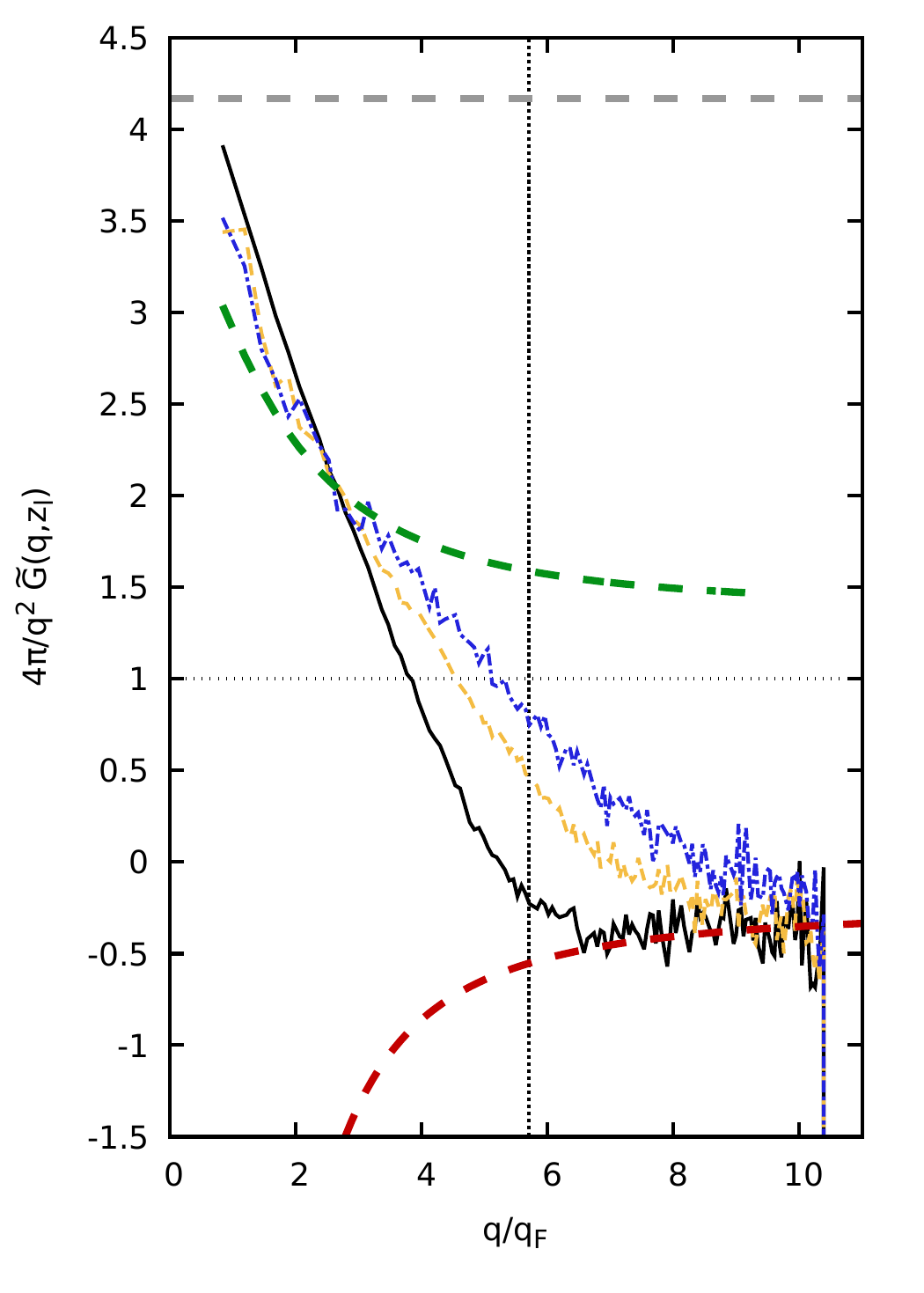}
\caption{\label{fig:rs2_theta2}
Left and right: dynamic Matsubara local field correction and re-scaled XC-kernel of the UEG at $r_s=2$ and $\Theta=2$
for $l=0$ (solid black), $l=1$ (double-dashed yellow), and $l=2$ (dash-dotted blue). Dashed red: Eq.~(\ref{eq:fit}) with $a$ being the single free fit parameter; dashed green: large-frequency limit $G(\mathbf{q},\infty)=\widetilde{G}(\mathbf{q},\mathrm{i}\infty)$~[Eq.~(\ref{eq:Ginfty})]; dashed grey: compressibility sum-rule describing the long wavelength limit $G(q\to0,0)=\widetilde{G}(q\to0,0)$. All limiting cases have been evaluated using the parametrization of the XC-free energy of the UEG by Groth \emph{et al.}~\cite{groth_prl} (GDSMFB).
The vertical dotted grey line in the right panel indicates the onset of the quantum tail in the momentum distribution $n(q)$~\cite{Hunger_PRE_2021}, cf.~Fig.~\ref{fig:Momentum_rs2_theta2}.
}
\end{figure}  

\begin{figure}\centering\includegraphics[width=0.45\textwidth]{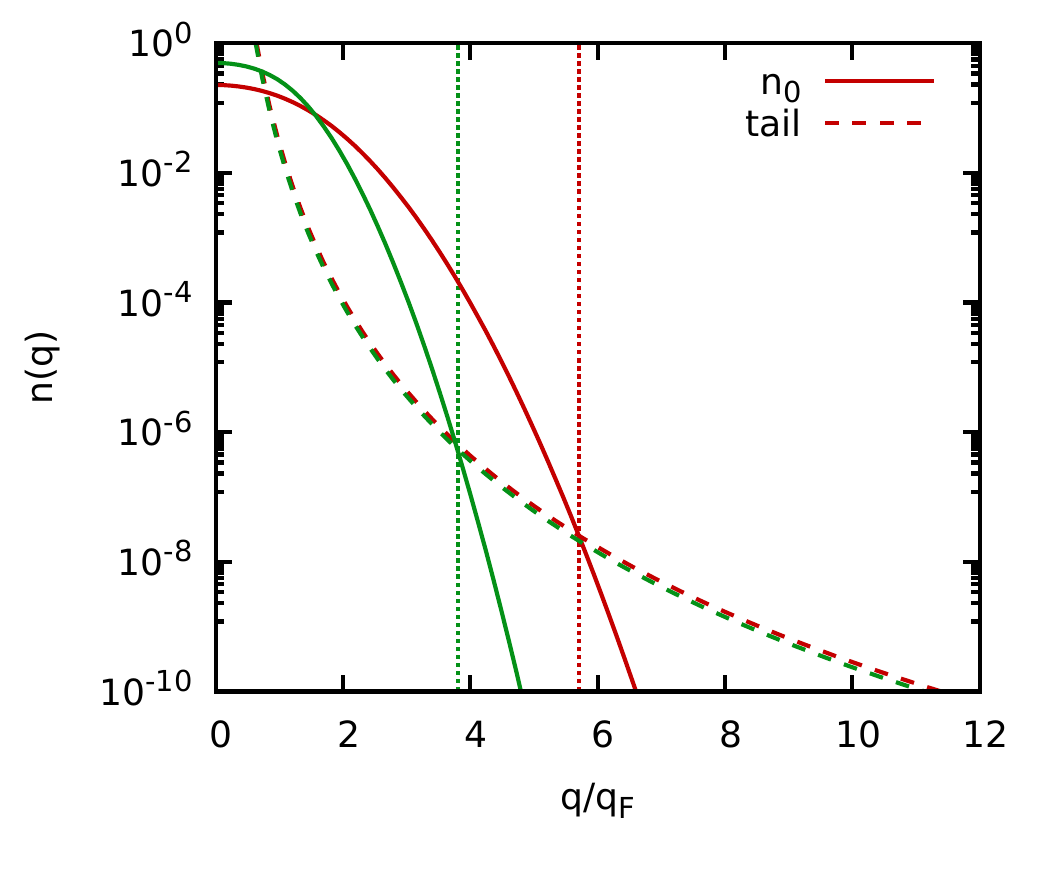}
\caption{\label{fig:Momentum_rs2_theta2}
Determining the onset of the quantum tail, Eq.~(\ref{eq:q_tail}), for $r_s=2$ and $\Theta=2$ [red] and $\Theta=1$ [green] following the procedure suggested by Hunger \emph{et al.}~\cite{Hunger_PRE_2021}.
}
\end{figure}

Let us begin our investigation with a study of the dynamic Matsubara local field correction in the warm dense matter regime. In the left panel of Fig.~\ref{fig:rs2_theta2}, we show new PIMC results for $\widetilde{G}(\mathbf{q},z_l)$ at $r_s=2$ and $\Theta=2$ for $N=14$ electrons. We note that the comparably small system size allows us to resolve $\widetilde{G}(\mathbf{q},z_l)$ up to ten times the Fermi wave number; finite-size effects are very minor for $q$-resolved properties at these conditions~\cite{dornheim_prl,review,Dornheim_JCP_2021,Holzmann_PRB_2016}, see Appendix A of Ref.~\cite{dornheim2024MatsubaraResponse}.
The solid black line shows the static limit of $\widetilde{G}(\mathbf{q},0)=G(\mathbf{q},0)$. In the long wavelength limit, it is governed by the well-known compressibity sum-rule~\cite{quantum_theory,Tolias_PRB_2024} that we evaluate from the accurate parametrization of the XC-free energy of the UEG by Groth \emph{et al.}~\cite{groth_prl} (GDSMFB), see the dashed grey line. With increasing $q$, the static LFC attains a maximum around $q\approx3.5q_\textnormal{F}$, followed by monotonic decrease; the slope appears to change around $q=6q_\textnormal{F}$, which can be seen particularly well in the right panel of Fig.~\ref{fig:rs2_theta2}, where we show the re-scaled XC-kernel, cf.~Eq.~(\ref{eq:kernel}).
The latter converges towards a constant for $q\gtrsim6q_\textnormal{F}$, as it is predicted by the analytical expression Eq.~(\ref{eq:Gzeroshort}) by Hou \emph{et al.}~\cite{Hou_PRB_2022}.
The dashed red lines show a fit of the LFC asymptote according to
\begin{eqnarray}\label{eq:fit}
    G(x) = a + bx^2
\end{eqnarray}
with $b$ following from Eq.~(\ref{eq:Gzeroshort}) and $a$ constituting the single fit parameter; the XC-contribution to the kinetic energy $T_\textnormal{xc}$ that enters $b$ has also been evaluated from the GDSMFB parametrization, and attains a negative value at these conditions~\cite{Militzer_Pollock_PRL_2002,Hunger_PRE_2021,Dornheim_PRB_nk_2021}. Hence, the local field correction diverges towards negative infinity in the limit of large $q$.
Evidently, the theoretically predicted asymptotic $q$-dependence fits well to the PIMC results.
An additional interesting point concerns the onset of the short wavelength limit with respect to the wavenumber. The vertical dotted grey line in the right panel of Fig.~\ref{fig:rs2_theta2} has been determined from the intersection of the ideal Fermi distribution function with the asymptotic tail of the momentum distribution function [Eq.~(\ref{eq:momentum_tail})], i.e., Eq.~(\ref{eq:q_tail}), which is illustrated by the red curves in Fig.~\ref{fig:Momentum_rs2_theta2}.
Evidently, this $q_\textnormal{tail}$ is in good qualitative agreement with the onset of the large-$q$ limit of the static LFC at the present conditions.


We next analyze the dynamic (i.e., $z_l$-dependent) behaviour of the LFC. The dashed green line in both panels of Fig.~\ref{fig:rs2_theta2} shows the high-frequency limit $\widetilde{G}(\mathbf{q},\mathrm{i}\infty)=G(\mathbf{q},\infty)$ [cf.~Eq.~(\ref{eq:Ginfty})] that has been discussed extensively e.g.~in the recent Ref.~\cite{dornheim2024MatsubaraResponse}. Since $T_\textnormal{XC}<0$, we find that $\widetilde{G}(\mathbf{q},\mathrm{i}\infty)$ increases with $q$, which is the opposite trend compared to the electron liquid regime at $r_s=20$ studied in Ref.~\cite{dornheim2024MatsubaraResponse}, and at $r_s=100$ studied in Sec.~\ref{sec:electron_liquid} below. This gives rise to the following interesting behavior in the WDM regime: for $q\lesssim3q_\textnormal{F}$, the dynamic Matsubara LFC converges towards its high-frequency limit from above; empirically, this is the case for all $q$ when $T_\textnormal{XC}>0$. For $q\gtrsim3q_\textnormal{F}$, however, $\widetilde{G}(\mathbf{q},z_l)$ converges towards $\widetilde{G}(\mathbf{q},\mathrm{i}\infty)$ from below, as the static and high-frequency limit diverge towards negative and positive infinity here. In addition, we find that the maximum in $\widetilde{G}(\mathbf{q},z_l)$ gets shifted to larger wavenumbers for increasing $l$. Finally, it appears that $\widetilde{G}(\mathbf{q},z_l)$ does indeed converge towards the same asymptotic short wavelength limit independent of $z_l$, but the onset wavenumber increases with $q$.
The asymptotic behaviour of the dynamic Matsubara LFC thus clearly depends on the sequence of the limits taken, and the double limit of infinite frequency and wavenumber does not exist.

\begin{figure}\centering\includegraphics[width=0.45\textwidth]{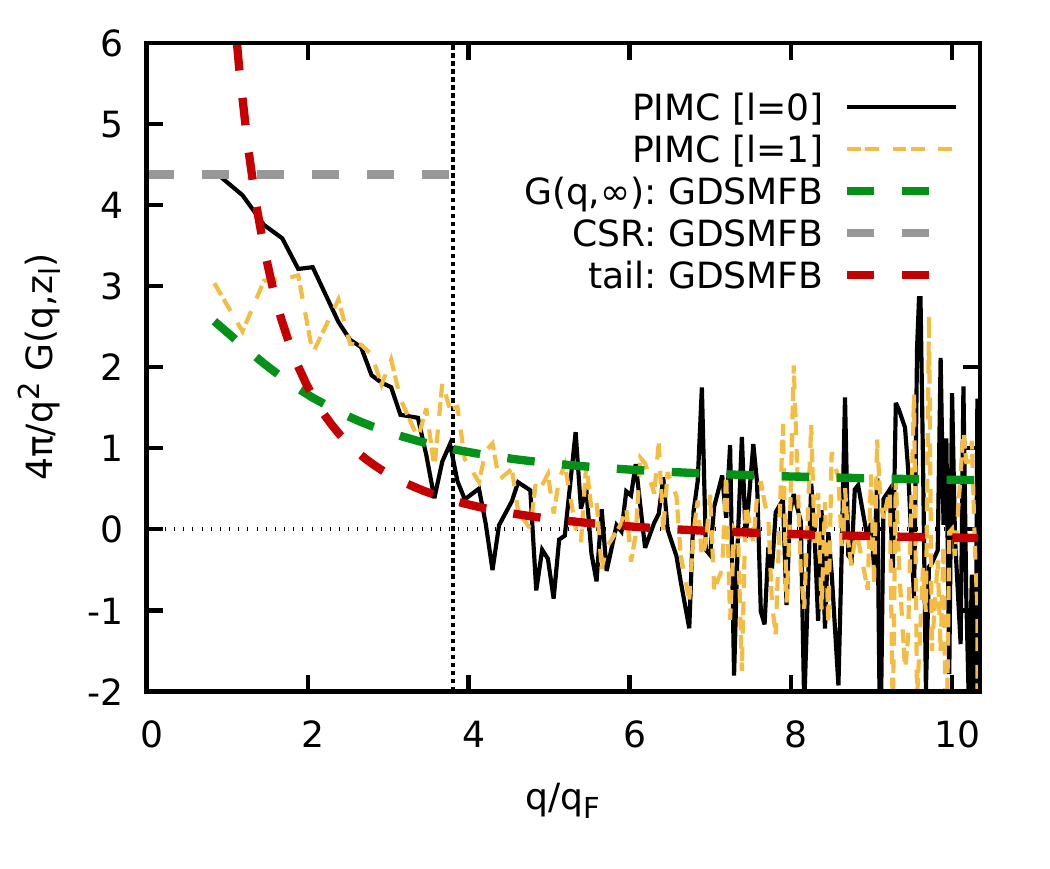}
\caption{\label{fig:rs2_theta1}
Rescaled dynamic Matsubara XC-kernel for $r_s=2$ and $\Theta=1$. The shift in the onset wavenumber of the short wavelength asymptote to smaller $q$ [vertical dotted grey line] compared to the case of $\Theta=2$ shown in Fig.~\ref{fig:rs2_theta2} is correctly predicted by the empirical criterion for the onset of the quantum tail in the momentum distribution $n(q)$ [Eq.~(\ref{eq:q_tail})], cf.~Fig.~\ref{fig:Momentum_rs2_theta2}.
}
\end{figure}  

We now consider the effect of temperature. In Fig.~\ref{fig:rs2_theta1}, we show the rescaled XC-kernel for $r_s=2$ and $\Theta=1$, i.e., at half the temperature compared to Fig.~\ref{fig:rs2_theta2}. Overall, we observe a qualitatively similar behavior to the latter case and restrict ourselves here to a report of the main differences. First, the larger fluctuations in the LFC are a direct consequence of the aforementioned fermion sign problem. Second, the density--temperature combination shown in Fig.~\ref{fig:rs2_theta1} has an almost vanishing value for $T_\textnormal{XC}$; consequently, the short wavelength limit of the dynamic Matsubara XC-kernel nearly vanishes. Still, it is $T_\textnormal{XC}<0$, leading to the same trends (i.e., convergence towards $\widetilde{G}(\mathbf{q},\mathrm{i}\infty)$ with $l$ from above for small $q$ and convergence from below for large $q$) as explained above.
Finally, we observe that the onset of the short wavelength asymptote in $\widetilde{G}(\mathbf{q},0)$ is shifted to a substantially smaller wavenumber for the lower temperature. This is in perfect agreement with the onset of quantum tail in $n(q)$, see the vertical dotted grey line in Fig.~\ref{fig:rs2_theta1} and the green data set in Fig.~\ref{fig:Momentum_rs2_theta2}.
Again, we find that $\widetilde{G}(\mathbf{q},z_l)$ attains the same $z_l\to\mathrm{i}\infty$ limit for $l\neq0$, although the statistical uncertainty increases with $l$.

\subsection{Electron liquid regime\label{sec:electron_liquid}}

\begin{figure}\centering\includegraphics[width=0.45\textwidth]{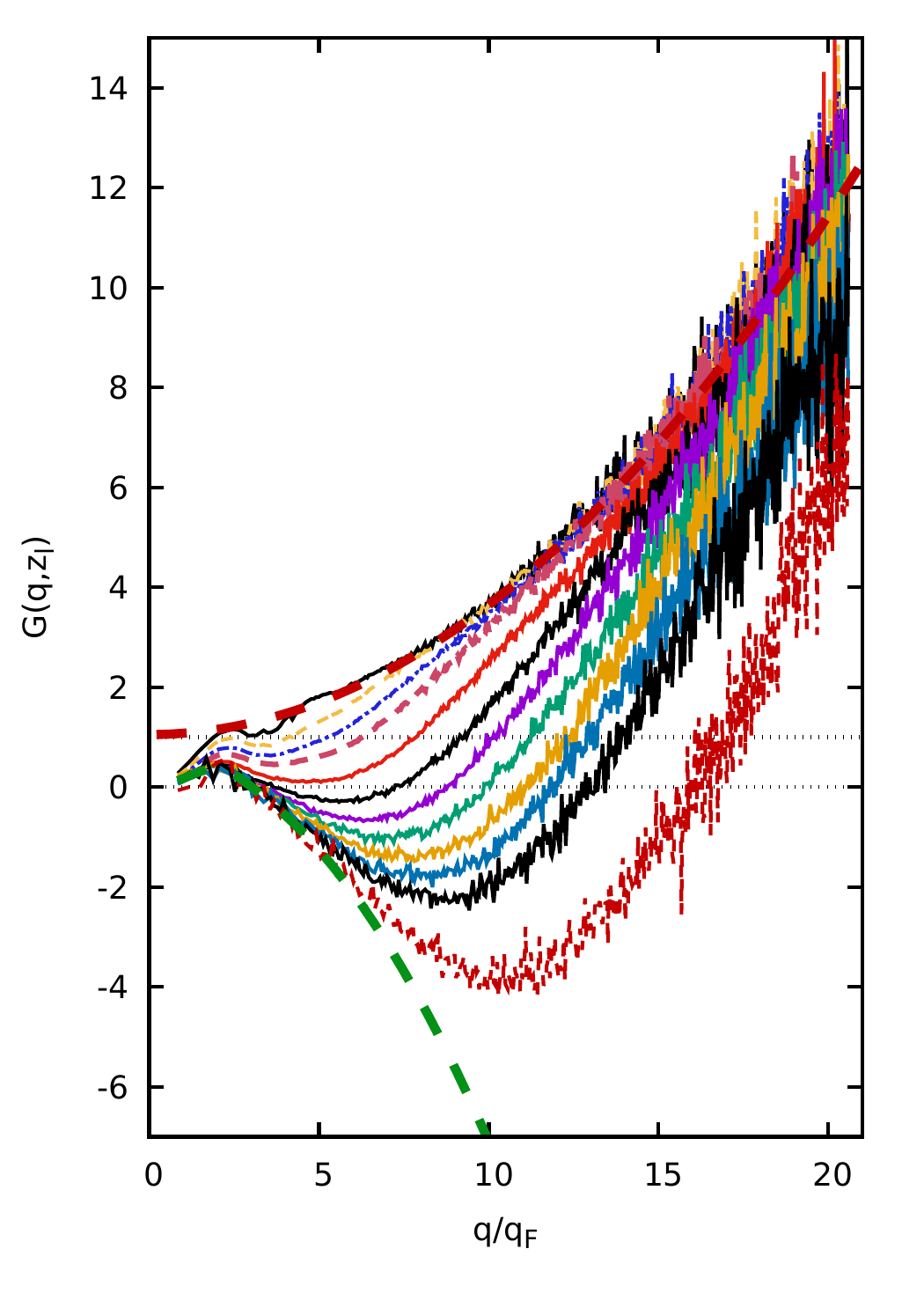}\includegraphics[width=0.45\textwidth]{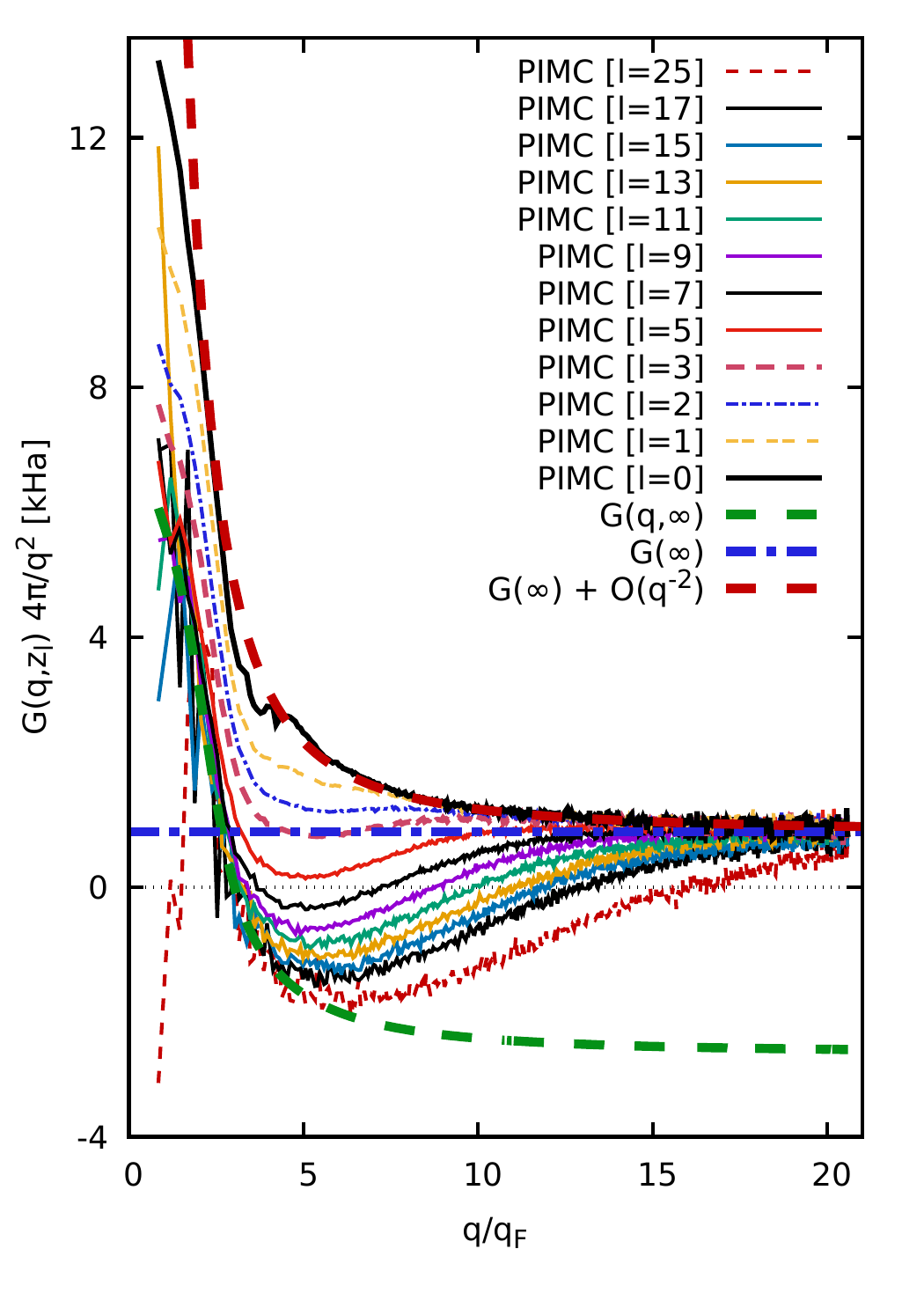}
\caption{\label{fig:full_rs100_theta1}
Left and right: dynamic Matsubara local field correction and re-scaled XC-kernel of the UEG at $r_s=100$ and $\Theta=1$
for various $z_l$ (see the legend). Bold dashed red: Eq.~(\ref{eq:fit}) with $a$ being the single free fit parameter; bold dashed blue: same as previous, without the $a$-term;
bold dashed green: large-frequency limit $G(\mathbf{q},\infty)=\widetilde{G}(\mathbf{q},\mathrm{i}\infty)$~[Eq.~(\ref{eq:Ginfty})].
}
\end{figure}  

Let us next focus on the strongly coupled electron liquid regime, which has been the focus of renewed recent interest~\cite{Takada_PRB_2016,Tolias_JCP_2021,dornheim_electron_liquid,Tolias_JCP_2023,Tolias_PRB_2024,koskelo2023shortrange,Dornheim_Nature_2022,dornheim2024MatsubaraResponse,Dornheim_EPL_2024,tanaka_hnc,Dutta_2013}.
While being out of reach for current experimental capabilities, these conditions allow one to focus on electron correlation effects, and to isolate their impact on observables such as the density response, which might be obscured by the interplay with other effects in the warm dense matter regime. Moreover, these conditions offer a challenging test bed for the development of novel theoretical approaches, most notably dielectric theories~\cite{tanaka_hnc,dornheim_electron_liquid,Tolias_JCP_2023,Tolias_PRB_2024,Tolias_JCP_2021,castello2021classical,arora}.
Finally, the electron liquid is interesting in its own right and gives rise to intriguing phenomena such as the \emph{roton} type feature~\cite{Trigger} in the dynamic structure factor~\cite{dornheim_dynamic,Dornheim_Nature_2022}, which has recently also been predicted to occur in warm dense hydrogen~\cite{Hamann_PRR_2023}.

In Fig.~\ref{fig:full_rs100_theta1}, we show the dynamic Matsubara LFC $\widetilde{G}(\mathbf{q},z_l)$ [left] and rescaled XC-kernel [right] for $r_s=100$ and $\Theta=1$. Due to the pronounced impact of electronic XC-effects onto the observables in our computer experiment---$F(\mathbf{q},\tau)$ and, via Eq.~(\ref{eq:MDR}) also $\widetilde{\chi}(\mathbf{q},z_l)$---we can resolve $\widetilde{G}(\mathbf{q},z_l)$ up to $q>20q_\textnormal{F}$, and for comparably large Matsubara frequencies; results for up to $l=25$ are shown in Fig.~\ref{fig:full_rs100_theta1}.
The static LFC exhibits a nontrivial structure with a clear local maximum around $q_\textnormal{max}\approx2q_\textnormal{F}$, followed by a second feature around $2q_\textnormal{max}$. This is a direct consequence of structural order and has been investigated in some detail in the recent Ref.~\cite{Dornheim_EPL_2024}.
The bold dashed green line shows the high-frequency limit $\widetilde{G}(\mathbf{q},\mathrm{i}\infty)=G(\mathbf{q},\infty)$. We note that, since neither parametrizations of the XC-free energy by Groth \emph{et al.}~\cite{groth_prl,review} nor by Karasiev \emph{et al.}~\cite{ksdt,status} extend to such low densities, we directly compute $T_\textnormal{XC}$ from our PIMC simulations. In particular, we find $T_\textnormal{XC}>0$ as the sign of the XC-contribution to the kinetic energy changes for $r_s\approx3$ at $\Theta=1$~\cite{Hunger_PRE_2021,Dornheim_PRB_nk_2021,Militzer_Pollock_PRL_2002}.
As it has been noted in earlier works~\cite{dornheim2024MatsubaraResponse,Dornheim_EPL_2024}, $\widetilde{G}(\mathbf{q},z_l)$ converges towards its high-frequency limit for increasingly large frequency with increasing the wavenumber due to the increased importance of quantum delocalization effects.

\begin{figure}\centering\includegraphics[width=0.45\textwidth]{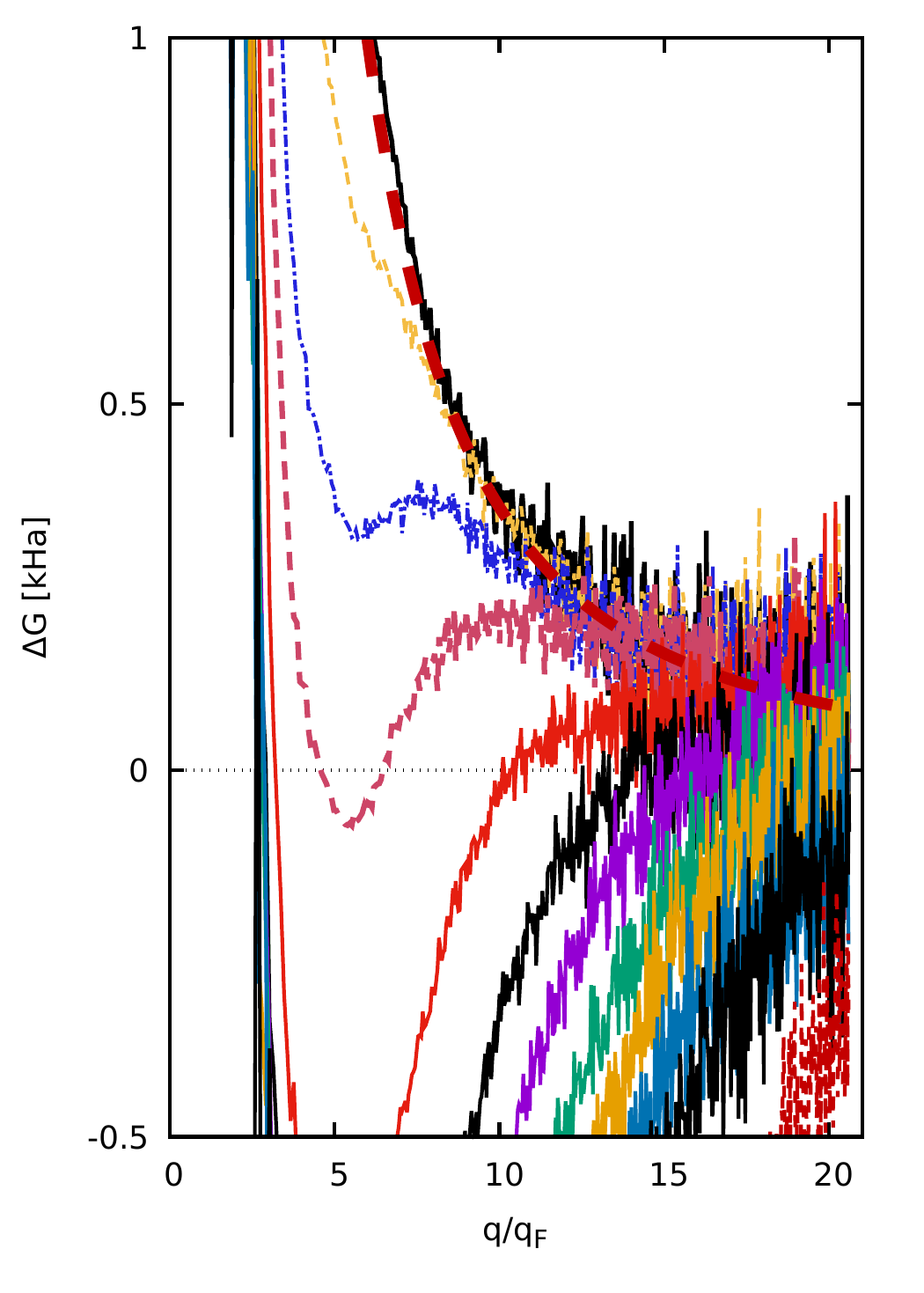}
\caption{\label{fig:I_am_the_limit_rs100_theta1}Convergence towards the short wavelength limit of the XC-kernel [cf.~Eq.~(\ref{eq:Gzeroshortgen})] for the data shown in Fig.~\ref{fig:full_rs100_theta1}, see Eq.~(\ref{eq:deviation_measure}).
}
\end{figure}  

We next come to the topic at hand, which is the short wavelength limit of $\widetilde{G}(\mathbf{q},z_l)$ in the electron liquid regime. From the right panel of Fig.~\ref{fig:full_rs100_theta1}, we find that the dynamic Matsubara LFC does indeed appear to converge against the same limit of Eq.~(\ref{eq:Gzeroshortgen}) [bold dash-dotted blue] independent of the Matsubara frequency $z_l$. However, while the curves for $l\lesssim4$ approach the limit from above, this seems to change for larger Matsubara frequencies.
In addition, the bold dashed red curve shows a fit to the static LFC (i.e., $l=0$) according to Eq.~(\ref{eq:fit}), with the constant $a$ being the only free parameter; it nicely fits to the asymptotic convergence towards the true short wavelength limit.

To get a more detailed view of the actual convergence of $\widetilde{G}(\mathbf{q},z_l)$ towards its asymptotic limit, we define a rescaled deviation measure
\begin{eqnarray}\label{eq:deviation_measure}
    \Delta \widetilde{G}(\mathbf{q},z_l) = \frac{4\pi}{q^2}\left[
\widetilde{G}(\mathbf{q},z_l) - G(\infty)
    \right]\ ,
\end{eqnarray}
where $G(\infty)$ is given by Eq.~(\ref{eq:Gzeroshortgen}). The results for Eq.~(\ref{eq:deviation_measure}) are shown in Fig.~\ref{fig:I_am_the_limit_rs100_theta1} and reveal a more nuanced behavior. While $\widetilde{G}(\mathbf{q},z_l)$ converges towards $G(\infty)$ monotonically from above for $l=0$ and $l=1$, there appear local minima around $q\approx6q_\textnormal{F}$ for $l\geq2$, which become more pronounced with increasing $l$.
Moreover, we observe that, at the present conditions, $\widetilde{G}(\mathbf{q},z_l)$ tends to first overshoot $G(\infty)$ from below, and then converges towards the short wavelength limit from above even for large $l$. Using the framework introduced by Tolias \emph{et al.}~\cite{tolias2024fouriermatsubara}, our PIMC simulations thus reveal the rich manifestation of dynamic XC-effects onto the linear density response of the strongly coupled electron liquid regime in the Matsubara frequency domain.


\section{Summary and Outlook\label{sec:summary}}

We have presented direct \emph{ab initio} PIMC results for the dynamic Matsubara local field correction $\widetilde{G}(\mathbf{q},z_l)$ of the finite-temperature uniform electron gas over an unprecedented range of wavenumbers $q$. This has allowed us to empirically confirm analytical results by Hou \emph{et al.}~\cite{Hou_PRB_2022}, and to complement their variational diagrammatic QMC results for $r_s=1$ by studying larger values of $r_s$.
First, we have verified the analytical short wavelength limit of the static local field correction $G(\mathbf{q},0)=\widetilde{G}(\mathbf{q},0)$. At the metallic density of $r_s=2$ in the warm dense matter regime, we have estimated $G(\mathbf{q},0)$ up to $q\sim10q_\textnormal{F}$; this is sufficient to resolve the onset of the short wavelength asymptote, which closely coincides with the empirical criterion for the onset of the quantum tail in the momentrum distribution function $n(\mathbf{q})$ suggested by Hunger \emph{et al.}~\cite{Hunger_PRE_2021}. In the strongly coupled electron liquid regime at $r_s=100$, we observe a number of nontrivial oscillations in the static LFC around integer multiples of the correlation peak around $q\approx2q_\textnormal{F}$ before $G(\mathbf{q},0)=\widetilde{G}(\mathbf{q},0)$ attains its short wavelength limit with the expected asymptotic behavior.
The second key result of the present work is given by the investigation of the fully dynamic Matsubara LFC $\widetilde{G}(\mathbf{q},z_l)$, which converges towards the same $q\to\infty$ limit independent of $z_l$, although the onset wavenumber of the asymptotic quantum tail systematically increases with $z_l$. At $r_s=100$, we have resolved $\widetilde{G}(\mathbf{q},z_l)$ up to $q\sim20q_\textnormal{F}$ and $l=25$, revealing a rich, non-monotonic and fairly nontrivial $q$-dependence at intermediate Matsubara frequencies. 

Our results further complete our understanding of the uniform electron gas, which constitutes the archetypical system of interacting electrons, and which is of deep relevance for statistical physics, quantum chemistry, and related fields. All PIMC results are freely available online~\cite{repo} and can be used to rigorously benchmark existing methods and guide the development of novel approaches, most notably in the field of dielectric theories. 
Moreover, the presented analysis of the short wavelength limit, and the empirical criterion for its onset wavenumber, constitute important ingredients for a future four-point parametrization of $\widetilde{G}(\mathbf{q},z_l;r_s,\Theta)$, which would be highly useful for a host of applications such as the construction of advanced XC-functionals for thermal DFT simulations.
Finally, we note that thoroughly understanding the imaginary-time dynamics of interacting quantum many-body systems is important in its own right. Indeed, there has been a remarkable recent focus on the ITCF $F(\mathbf{q},\tau)$ of warm dense matter systems, both from a theoretical~\cite{Dornheim_insight_2022,Dornheim_PTR_2022,dornheim2024ab,Dornheim_review}
and an experimental perspective~\cite{Dornheim_T_2022,Dornheim_T2_2022,dornheim2023xray,Schoerner_PRE_2023}. Two particularly intriguing projects for impactful future research are thus given by the analysis of $\widetilde{\chi}(\mathbf{q},z_l)$ and $\widetilde{G}(\mathbf{q},z_l)$ of real materials such as warm dense hydrogen and other light elements based on highly accurate PIMC simulations, and the analysis of experimental observations in terms of the dynamic Matsubara density response based on X-ray Thomson scattering results for $F(\mathbf{q},\tau)$.







\section*{Acknowledgments}
This work was partially supported by the Center for Advanced Systems Understanding (CASUS), financed by Germany’s Federal Ministry of Education and Research (BMBF) and the Saxon state government out of the State budget approved by the Saxon State Parliament. Further support is acknowledged for the CASUS Open Project \emph{Guiding dielectric theories with ab initio quantum Monte Carlo simulations: from the strongly coupled electron liquid to warm dense matter}. This work has received funding from the European Research Council (ERC) under the European Union’s Horizon 2022 research and innovation programme
(Grant agreement No. 101076233, "PREXTREME"). 
Views and opinions expressed are however those of the authors only and do not necessarily reflect those of the European Union or the European Research Council Executive Agency. Neither the European Union nor the granting authority can be held responsible for them. Computations were performed on a Bull Cluster at the Center for Information Services and High-Performance Computing (ZIH) at Technische Universit\"at Dresden and at the Norddeutscher Verbund f\"ur Hoch- und H\"ochstleistungsrechnen (HLRN) under grant mvp00024. 

\section*{References}

\bibliographystyle{unsrt}
\bibliography{bibliography}

\begin{thebibliography}{100}
\bibitem{repo}
A link to a repository containing all PIMC raw data will be made available upon publication.
\end{thebibliography}

\end{document}